\documentclass[a4paper,12pt]{article}
\usepackage{fullpage}
\usepackage{amsmath}
\usepackage{amsfonts}
\usepackage{graphicx}

\newcommand{\be}{\begin{equation}}
\newcommand{\ee}{\end{equation}}
\def\bea{\begin{eqnarray}}
\def\eea{\end{eqnarray}}

\def\lsim{\mathrel{\rlap{\lower4pt\hbox{\hskip1pt$\sim$}}\raise1pt\hbox{$<$}}}                
\def\gsim{\mathrel{\rlap{\lower4pt\hbox{\hskip1pt$\sim$}}\raise1pt\hbox{$>$}}}                

\def\={\! =\! }
\def\-{\! -\! }
\def\+{\! +\! }
\def\lb{\left( }
\def\rb{\right) }
\def\d{\partial}
\newcommand{\bphi}{\mbox{\boldmath $\phi$}}

\newcommand{\bff}{\mbox{\boldmath $f$}}

\newcommand{\bF}{\mbox{\boldmath $F$}}
\newcommand{\bolde}{\mbox{\boldmath $e$}}

\newcommand\T{\rule{0pt}{2.6ex}}
\newcommand\B{\rule[-1.2ex]{0pt}{0pt}}

\title{Hopf Solitons in the AFZ Model}
\author{Mike Gillard\\
{\em \normalsize Department of Mathematical Sciences,}\\
{\em \normalsize Durham University, Durham DH1 3LE, U.K.}\\
{\normalsize Email: mike.gillard@durham.ac.uk}
}

\begin{document}
\date{}
\maketitle
\begin{abstract}
The Aratyn-Ferreira-Zimerman (AFZ) model is a conformal field theory in three-dimensional space. It has solutions that are topological solitons classified by an integer-valued Hopf index. There exist infinitely many axial solutions which have been found analytically. Axial, knot and linked solitons are found numerically to be static solutions using a modified volume preserving flow for Hopf index one to eight, allowing for comparison with other Hopf soliton models. Solutions include a static trefoil knot at Hopf index five.  A one-parameter family of conformal Skyrme-Faddeev (CSF) models, consisting of linear combinations of the Nicole and AFZ models, are also investigated numerically. The transition of solutions for Hopf index four is mapped across these models. A topological change between linked and axial solutions occurs, with fewer models permitting axial solitons than linked solitons at Hopf index four.       
\end{abstract}

\section{Introduction}

The Skyrme-Faddeev model \cite{skyrmefad}\cite{fadniemi} is the most well known model which has Hopf solitions. The model is of physical interest in both quantum field theory  and condensed matter physics. It has been used as a possible description of potential knotted solitons in multicomponent superconductors \cite{matter1}\cite{matter2} and it is of potential relevance in a low energy effective theory of QCD \cite{yang1}\cite{yang2}\cite{yang3}. It has a Lagrangian density containing two parts, a sigma-type term and a Skyrme-type term, where $\bphi\=(\phi_1,\phi_2,\phi_3)$ maps to the two-sphere:
\be
\mathcal L_{FN}=\lb\d^\mu\bphi\cdot\d_\mu\bphi\rb-\kappa\lb\d^\mu\bphi\times\d^\nu\bphi\rb\cdot\lb\d_\mu\bphi\times\d_\nu\bphi\rb.
\ee
In $3+1$ dimensions, neither term on its own will allow for the possibility of solitons by Derricks theorem \cite{derrick}, but the combination of the terms does. Taking either term to a specific fractional power may also allow for the possibility of topological solitons. Taking the sigma-term to the power three half's is referred to as the Nicole model \cite{nicole},
\be
\mathcal L_{Ni}=\lb\d^\mu\bphi\cdot\d_\mu\bphi\rb^{\frac{3}{2}}.
\label{LNi}
\ee
The simplest Hopf map is a solution to the Nicole model, and is the only analytic solution known. Further axial configurations for higher Hopf charges have been found by imposing axial symmetry \cite{nic06}. Non-axial solitons in this models have already been investigated \cite{nicvp}. Taking the Skyrme-term to the power three quarters is referred to as the AFZ model \cite{AFZ1}\cite{AFZ2}, after Aratyn, Ferreira and Zimerman who found infinitely many analytic solutions using toroidal coordinates. The Lagrangian density is given by
\be
\mathcal L_{AFZ}=\lb H^2_{\mu\nu}\rb^{\frac{3}{4}}\, , \quad \text{where} \quad H_{\mu\nu}=\bphi\cdot\lb\d_\mu\bphi\times\d_\nu\bphi\rb.
\label{LAFZ}
\ee
In these models the target space is an $\mathcal S^2$, requiring that $\bphi\cdot\bphi\=1$. The fractional powers for the Nicole and AFZ models are chosen to give scale invariant field theories. 

Each of the above describe a map $\bphi:\mathbb R^3\to\mathcal S^2$, with $\mathbb R^3$ being topologically equivalent to $\mathcal S^3$, if the field is single-valued at spacial infinity. This is also required for finite energy in the Skyrme-Faddeev and Nicole models, and is taken to be $\bphi(\infty)\=(0,0,1)\equiv\bolde_3$. Solitons in these theories will then be classified by an integer valued topological charge $\mathcal Q$, given by the class of smooth maps $\bphi:\mathcal S^3\to\mathcal S^2$, i.e.\ $\pi_3(\mathcal S^2)\=\mathcal Z$. This charge is topological, meaning that maps of different charges cannot be continuously deformed into each other. This charge is referred to as the Hopf index. Pre-images of points of the target $\mathcal S^2$ are generally closed curves in $\mathbb R^3$. The Hopf index counts the number of times that the pre-image curves of two separate points are linked together. Continuous deformations of the inverse map must leave the linking unchanged, so each Hopf index splits the set of all maps, $\bphi$, into distinct sets. This means that any static solutions will have a given topological charge, $\mathcal Q$, and the static solution with the lowest energy for a given $\mathcal Q$ will be a stable soliton.  

The purpose of this paper will be to find the minimal energy solutions for Hopf charges one to eight using numerical simulations of the full nonlinear field theory. Minimisation will utilise volume preserving flow \cite{nicvp}\cite{volpres} to overcome technical difficulties relating to scale invariant field theories and simulations will not be restricted by symmetry. Further technical difficulties appear in the AFZ model resulting from an infinite-dimensional symmetry group acting on target space \cite{AFZsym}\cite{AFZsym3}. The AFZ model has area preserving diffeomorphisms of the target $\mathcal S^2$ that are symmetries of the equations of motion. Combining these with the conformal base space symmetry results in a Noether charge which has a divergent flux. Examples of this flux diverging along an infinite axis \cite{AFZsym} and along a ring \cite{AFZsym4} have been found.  Sectors of the model with different Hopf index are also not separated by an infinite energy barrier, where a symmetry transformation relates all solutions to the vacuum \cite{AFZsym2}. Solutions to the AFZ model have been found \cite{AFZ2} so this problem can be overcome (see \cite{AFZsym2} for details). Fixing the value of $\bphi(\infty)$ will do this. The Nicole model does not have the same target space symmetries \cite{AFZsym5}, so adding a small amount of the Nicole model to the minimisation of the AFZ model will break the target space symmetries. This will be the approach used below, where a contribution of around $1\%$ of the energy from the Nicole model will be sufficient. 

The addition of the Nicole symmetry breaking term to the minimisation motivates investigation of a one-parameter family of conformal Skyrme-Faddeev (CSF) models. The set consists of all linear combinations of the Nicole and AFZ models where the coefficients sum to one. The transition of the energy and topological changes of solutions to these models will be simulated numerically using volume preserving flow. The investigation will concentrate on a Hopf index which has topologically different solutions in the Nicole and AFZ models. Details of these models will follow the discussion of the AFZ model. 

\section{AFZ Model}\label{AFZsection}

The interest here will be finding static solutions of the AFZ model, and so this system is best described by its static energy,
\be
E_{AFZ}=\dfrac{1}{16\pi^2 2^\frac{3}{4}}\int \lb H_{ij}^2\rb^{\frac{3}{4}}d^3x\, , \quad H_{ij}=\bphi\cdot\lb \d_i\bphi\times\d_j\bphi\rb,
\label{EAFZ}
\ee
with spacial indices, $i,j\=1\,.\,.\,3$. The normalisation gives the charge $\mathcal Q\= 1$ soliton an energy, $E_{AFZ}\=1$. The position of the solitons will be defined where the field is antipodal to the vacuum, i.e.\ $\bphi\=\-\bolde_3$. These will be closed string(s) in $\mathbb R^3$. 

It is possible to find an infinite number of solutions to the AFZ model by exploiting its symmetries in toroidal coordinates \cite{AFZ2}. This leads to an energy that depends only on the windings of the soliton about the two cycles of the torus, $m$ and $n$. The axial solitons are denoted $\mathcal A_{n,m}$ with Hopf index $\mathcal Q\=nm$. The solutions are then dependent on a profile function, which can be found explicitly for any axial soliton. The energy is 
\be
E_{axial}=2^{-\frac{1}{2}}\sqrt{|n||m|\lb|n|+|m|\rb}.
\label{AFZaxial}
\ee
This can be re-expressed using a ratio of these winding numbers to show that the energy of an axial soliton, for a given $\mathcal Q$, is minimised when the two winding numbers are as close as possible to one another. 
\be
E_{axial}=2^{-\frac{1}{2}}\mathcal Q^\frac{3}{4}\sqrt{\vartheta+\frac{1}{\vartheta}}\,, \qquad \vartheta^2=\left|\dfrac{m}{n}\right|.
\label{AFZaxialQ}
\ee

The Skyrme-Faddeev model has an energy lower bound of the form $E_{SF}\ge c_{SF}\mathcal Q^\frac{3}{4}$, where $c_{SF}$ is a known constant \cite{vakkap}. No proof for a lower bound of this type exists for either the Nicole or AFZ model. If such a bound exists then in the Nicole model the maximum possible value for the constant is $c_{Ni}\=1$ from the exact $\mathcal Q\=1$ solution, where the energy is normalised to $1$ \cite{nicole}. The work in \cite{nicvp} suggests solitons for $\mathcal Q>2$ have an energy $\sim10\-12\%$ above this conjectured bound. In the AFZ model the largest possible value for the constant is also $c_{AFZ}\=1$ from (\ref{AFZaxialQ}) \cite{AFZ2}. This bound is attained for all square charges $\mathcal Q\=n^2$. Clearly, axial solutions when the Hopf index is a prime number will fall well above this bound. In fact, static solutions to the Nicole model for Hopf index five and seven have an energy in the AFZ model, $E_{AFZ}\sim 10\-15\%$ under the energy of the axial solutions (\ref{AFZaxialQ}). Although these are not static solutions for the AFZ model, it does suggest static non-axial solitons must exist for some $\mathcal Q$.

\subsection{Volume Preserving Flow}

The local energy density $\mathcal E_{AFZ}$ is given by 
\be
\mathcal E_{AFZ} = H^{\frac{3}{4}}+\lambda_{AFZ}\lb 1-\bphi\cdot\bphi\rb, \quad \text{where} \quad H_{ij}^2\equiv H.
\ee
The Lagrange multiplier, $\lambda_{AFZ}$, constrains the system to the $\mathcal S^2$. Variation of the energy density leads to the static equations of motion for the AFZ model,
\bea
\label{eom}
& & H^{-\frac{5}{4}}\left[ \lb\d_i\d_j\bphi\cdot\d_j\bphi\rb\Lambda-\lb\d_i\d_j\bphi\cdot\d_k\bphi\rb\lb\d_j\bphi\cdot\d_k\bphi\rb\right]\left[\d_l\bphi\lb\d_i\bphi\cdot\d_l\bphi\rb\-\d_i\bphi\Lambda\right]\\
& & +H^{-\frac{1}{4}}\left[\nabla^2\bphi\Lambda+\d_i\bphi\lb\d_i\d_j\bphi\cdot\d_j\bphi\rb\-\d_i\d_j\bphi\lb\d_i\bphi\cdot\d_j\bphi\rb\-\d_j\bphi\lb\nabla^2\bphi\cdot\d_j\bphi\rb\right]+\lambda_{AFZ}\bphi=\mbox{\boldmath $0$},\nonumber
\eea
where $\Lambda\=\lb\d_i\bphi\cdot\d_i\bphi\rb$ and $\nabla^2\bphi\=\d_i\d_i\bphi$. $\lambda_{AFZ}$ is obtained from (\ref{eom}), using $\bphi\cdot\bphi\=1$. The gradient flow equations, with flow time parameter, $t'$, are
\be
\dfrac{\d\bphi}{\d t'}=-\dfrac{\delta\mathcal E_{AFZ}}{\delta\bphi}\equiv \bF'=\bF_{AFZ}+\lambda_{AFZ}\bphi.
\ee
They give the direction of steepest decent of the energy functional (\ref{EAFZ}), where
\bea
\label{FAFZ}
\bF_{AFZ} & \!\!\! = &\!\!\! H^{-\frac{1}{4}}\left[\nabla^2\bphi\Lambda+\d_i\bphi\lb\d_i\d_j\bphi\cdot\d_j\bphi\rb\-\d_i\d_j\bphi\lb\d_i\bphi\cdot\d_j\bphi\rb\-\d_j\bphi\lb\nabla^2\bphi\cdot\d_j\bphi\rb\right]\\
& & \!\!\!\!\!\!\!\!\!
+H^{-\frac{5}{4}}\left[\lb\d_i\d_j\bphi\cdot\d_j\bphi\rb\Lambda\-\lb\d_i\d_j\bphi\cdot\d_k\bphi\rb\lb\d_j\bphi\cdot\d_k\bphi\rb\right]\left[\d_l\bphi\lb\d_i\bphi\cdot\d_l\bphi\rb\-\d_i\bphi\Lambda\right],\nonumber
\eea
and $\bphi\cdot\bF'\=0$ is needed to keep the flow on the unit sphere, so $\lambda_{AFZ}\=-\bphi\cdot\bF_{AFZ}$. 

An additional symmetry breaking term can to be added to the minimisation to overcome technical issues in the AFZ model. The local energy density becomes 
\be
\mathcal E = H^{\frac{3}{4}}+\epsilon\Lambda^\frac{3}{2}+\lambda\lb 1-\bphi\cdot\bphi\rb,
\ee
again using a Lagrange multiplier $\lambda$ and $\epsilon$ is a constant that weights the impact of the Nicole model on the flow. Since energies in the Nicole model and AFZ model are of the same order, $\epsilon\= 0.01$ would add $\sim1\%$ of Nicole model to the overall flow direction. The new direction of flow, $\bF$, will be given by $\bF\=\bF_{AFZ}+\epsilon\bF_{Ni}+\lambda\bphi$, where the gradient flow of the AFZ model, $\bF_{AFZ}$, is the same as (\ref{FAFZ}) and the gradient flow of the Nicole model, $\bF_{Ni}$, is 
\be
\bF_{Ni}=\nabla^2\bphi\Lambda^\frac{1}{2}+\d_i\bphi\lb\d_i\d_j\bphi\cdot\d_j\bphi\rb\Lambda^{-\frac{1}{2}}.
\label{FNi}
\ee

Minimisation using the gradient flow, $\bF$, will not find stable solutions. Zero modes associated with changes of scale will be broken when the theory is placed on a lattice. These modes will cause the solitons to shrink until their scale becomes too small and the solutions fall through the lattice. This behaviour is seen in the O$(3)$ sigma model \cite{O3shrink} and can overcome using a novel lattice configuration \cite{wardlattice}, but no such lattice configuration is known here. This issue has already been overcome for the Nicole model using a modified gradient flow that preserves the size of the soliton \cite{nicvp}. This volume preserving flow \cite{volpres}\cite{nicvp} can be used to constrain the minimisation to a direction which will leave a given total integral of a given function, $v$,  unchanged. This flow is given by
\be
\dfrac{\d\bphi}{\d t}= \bF-\dfrac{\left<\bff\cdot\bF\right>}{\left<\bff\cdot\bff\right>}\bff,\quad \text{with} \quad \bff=-\dfrac{\delta v}{\delta\bphi}, \quad \text{and} \quad v=\dfrac{1}{2}\lb1-\phi_3\rb.\label{volpresflowafz}
\ee
This particular flow will preserve, in general, the size of the soliton. Since the soliton position is defined to be where $\phi_3\=-1$ and $\phi_3\!\to\! 1$ at spacial infinity, this choice of volume counting seems reasonable. This flow will reduce the energy, $E$, and keep the volume, $V$, constant. A simple adaptation of the proof in \cite{volpres} will show this, but both should be clear from the construction of the volume preserving flow.  

Volume preserving flow will be applied to configurations of Hopf index $\mathcal Q\=1\,.\,.\,8$ using initial conditions generated by rational maps \cite{knots}, details of which will follow shortly. These configurations will exist in a finite region, $\Omega$, and the field will be fixed at $\bphi\=\lb 0,0,1\rb$ on $\d\Omega$. For the purpose of numerical investigation $\Omega$ will be a cubic lattice, with unit spacing, of length $150$. Derivatives will be approximated to fourth order accuracy using a finite difference method and the fields evolved using an explicit method with timestep $\Delta t\le0.1$. The AFZ energy and volume for configurations will be
\be
E_{AFZ}=\dfrac{1}{16\pi^2 2^\frac{3}{4}}\int_\Omega H^\frac{3}{4}\, d^3x\, , \qquad V=\int_\Omega v\, d^3x.
\ee
These integrals and inner products will be evaluated as summations over the lattice. The timestep will be reduced, where appropriate, to allow for continued evolution for a given $\epsilon$ which may be necessary if symmetry issues of the AFZ model become problematic. This will be done to keep $\epsilon$ as small as possible. Simulations will use $\epsilon\=0.01$ in general, but $\epsilon\lsim 0.05$ may be used for small periods of evolution to circumvent symmetry issues which cannot be overcome efficiently by lowering the timestep. In all cases except for the charge 1 soliton $\epsilon\! >\! 0$ is needed, although minimising a charge 1 soliton with $\epsilon\! >\! 0$ reaches almost the same static solution as without the symmetry breaking. 

\subsection{Initial Conditions}

Initial conditions will be generated using rational maps \cite{knots}. The fields are constructed by mapping $(x_1,x_2,x_3)\in\mathbb R^3$ to the unit three-sphere using a degree one spherically equivariant map. In terms of complex coordinates on the three-sphere, $Z_1,Z_0$, with $|Z_1|^2\+|Z_0|^2\=1$,
\be
\lb Z_1,Z_0\rb=\lb\lb x_1+ix_2\rb\dfrac{\sin{f}}{r},\cos{f}+ix_3\dfrac{\sin{f}}{r}\rb.
\ee 
Here, $f(r)$ is a monotonically decreasing function of the radius, $r$. It has boundary conditions $f(0)\=\pi$ and $f(\infty)\=0$. As our solutions will exist in a finite region, $\Omega$, $f\=0$ on $\d\Omega$. Initial conditions are then generated when the stereographic projection of $\bphi$ is a rational function of $Z_1$ and $Z_0$, such that
\be
W=\dfrac{\phi_1+i\phi_2}{1+\phi_3}=\dfrac{p\lb Z_1,Z_0\rb}{q\lb Z_1,Z_0\rb},
\label{W}
\ee
where $p$ and $q$ are polynomials in $Z_1$ and $Z_0$, and $W$ is the Riemann sphere coordinate. 

Axial solitons can be found by taking $p(Z_1)$ and $q(Z_0)$ such that (\ref{W}) becomes
\be
W=\dfrac{Z_1^n}{Z_0^m}.
\ee
These are $\mathcal A_{n,m}$ type fields with a Hopf index $\mathcal Q\=nm$. 

Knotted fields can also be generated from (\ref{W}) when
\be
W=\dfrac{Z_1^\alpha Z_0^\beta}{Z_1^a+Z_0^b}.
\ee 
These are torus knots denoted $\mathcal K_{a,b}$ and have Hopf index $\mathcal Q\=\alpha b\+\beta a$. Here, $a>b$ and are coprime positive integers with $\alpha>0$ and $\beta\ge0$. The only knotted solution of interest here will be the trefoil knot, $\mathcal K_{3,2}$.

Linked configurations can be generated from rational maps with reducible denominators, $q$. Two-component links are denoted $\mathcal L^{a,b}_{n,m}$. The integers $n,m$ represent the Hopf charges of the individual components and $a,b$ the number of times each component links with the other. They will have a Hopf index $\mathcal Q\=a\+b\+n\+m$. On such example is
\be
W=\dfrac{Z_1^{n+1}}{Z_1^2-Z_0^2}=\dfrac{Z_1^n}{2\lb Z_1-Z_0\rb}+\dfrac{Z_1^n}{2\lb Z_1+Z_0\rb}.
\ee  
This is an $\mathcal L^{1,1}_{n,n}$ link with Hopf index $\mathcal Q\=2n\+2$ consisting of two $\mathcal A_{n,1}$'s linked once each. Other links can be produced from similar rational maps. 

See \cite{knots} for further details of Hopf soliton rational maps. 

\subsection{Solitons in the AFZ model}

\begin{table}[t!]
\vspace{-10pt}
 \begin{center}
\begin{tabular}{  | r | c | c | c | c | r | c | c | c | c | }
\hline
\multicolumn{10}{|c|}{\large{\textbf{AFZ Energy \T \B}}} \\ 
\hline	
\multicolumn{4}{|c|}{\textbf{Axial Solutions}\T \B} &  & \multicolumn{5}{|c|}{\textbf{Numerical Results}} \\ 
\hline	
$\mathcal{Q}$ & \textbf{Type} & $E$  &  $E/\mathcal Q^\frac{3}{4}\T \B$  & & $\mathcal{Q}$ & \textbf{Type} & $\epsilon$  & $E$ & $E/\mathcal Q^\frac{3}{4}$ \\ \hline\hline
   1 \T & $A_{1,1}$ & 1.000 & 1.000 & & 1 & $A_{1,1}$         & 0.01 & 1.003 & 1.003 \\ \hline
   2 \T & $A_{2,1}$ & 1.732 & 1.030 & & 2 & $A_{2,1}$         & 0.01 & 1.738 & 1.033 \\ \hline
   3 \T & $A_{3,1}$ & 2.449 & 1.075 & & 3 & $\tilde{A}_{3,1}$ & 0.01 & 2.414 & 1.059 \\ \hline
   4 \T & $A_{2,2}$ & 2.828 & 1.000 & & 4 & $A_{2,2}$         & 0.01 & 2.852 & 1.008 \\ \hline
   4 \T & $A_{4,1}$ & 3.162 & 1.118 & & 5 & $K_{3,2}$         & 0.01 & 3.391 & 1.014 \\ \hline 
   5 \T & $A_{5,1}$ & 3.873 & 1.158 & & 5 & $L^{1,1}_{1,2}$   & 0.01 & 3.451 & 1.032 \\ \hline
   6 \T & $A_{3,2}$ & 3.873 & 1.010 & & 6 & $A_{3,2}$         & 0.01 & 3.907 & 1.019 \\ \hline
   7 \T & $A_{7,1}$ & 5.292 & 1.230 & & 7 & $K_{3,2}$         & 0.01 & 4.407 & 1.024 \\ \hline
   8 \T & $A_{4,2}$ & 4.899 & 1.030 & & 8 & $A_{4,2}$         & 0.01 & 4.954 & 1.041 \\ 
    \hline
  \end{tabular}
\end{center}
\vspace{-15pt}
\caption{\footnotesize{(left) List of AFZ axial soliton energies, calculated using (\ref{AFZaxial}) and (right) are the energies of the static solutions found using the volume preserving gradient flow method. The amount of Nicole model used to find the solutions, $\epsilon$ is also given.}}
\label{AFZtable}
\vspace{-10pt}
\end{table}
\begin{figure}[t]
\vspace{-10pt}
\begin{center}
\includegraphics[scale=1.05]{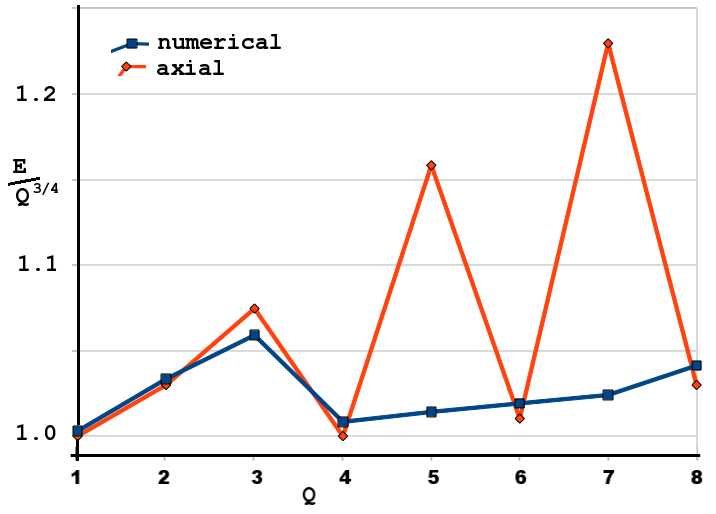}
\end{center}
\vspace{-20pt}
\caption{\footnotesize{Plot of $E/\mathcal Q^\frac{3}{4}$ against charge from table \ref{AFZtable} for both the exact axial solution (\ref{AFZaxial}) (\textit{red}), and the numerically calculated energies (\textit{blue}).}}
\label{fig-afzgraph}
\vspace{-10pt}
\end{figure}

The previously defined rational maps create field configurations which have a given Hopf index. By minimising the energy of different rational maps, potential local and global energy minima may be found for a variety of Hopf charges in the AFZ model. This section will classify those results, as well as noting transitions between different types of configuration. 

\begin{figure}[t]
\centering
\vspace{-10pt}
\includegraphics[scale=0.7]{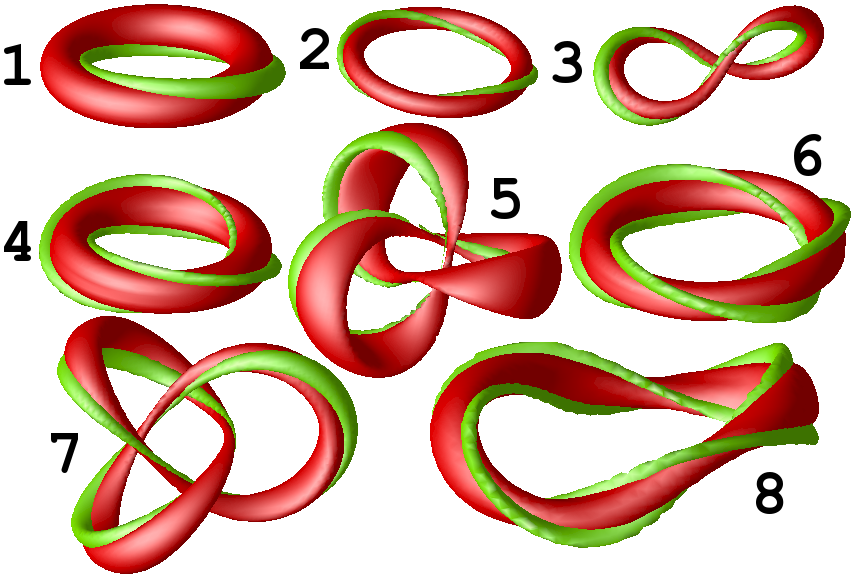}
\vspace{-10pt}
\caption{\footnotesize{Hopf index $\mathcal Q\!=\!1\!-\!8$ minimal energy soliton solutions in the AFZ model. The numbers indicate the Hopf index $\mathcal Q$. 
}}
\label{fig-afzmin}
\vspace{-10pt}
\end{figure}

The soliton energies for both the axial solutions and numerical results are given in table \ref{AFZtable} and are plotted in figure \ref{fig-afzgraph}. The minimal energy soliton positions and linking are shown in figure \ref{fig-afzmin}. All local minima, as well as details on initial conditions, will be discussed. The soliton plots (figure \ref{fig-afzmin}) show a level set of $\phi_3\=-0.9$, giving an idea of the position of the soliton and a level set $\bphi\=(\sqrt{\mu\lb2-\mu\rb}, 0, \mu-1)$, with $\mu=0.1$ which shows the linking. This is an arbitrary choice of a second pre-image curve used to see the linking number, and so the Hopf index, of a given configuration. 

Hopf index one and two solitons are the same as in the Skyrme-Faddeev and Nicole models. They are the axial $\mathcal A_{1,1}$ and $\mathcal A_{2,1}$ configurations, see figure \ref{fig-afzmin}. The $\mathcal A_{1,1}$ axial soliton has an energy of $1.003$ for a volume $V\=115000$. This matches well with the exact axial solution and is $<0.5\%$ above the expected energy. The $\mathcal A_{2,1}$ axial soliton has an energy of $1.738$ with $V\=100000$. This is also $<0.5\%$ above the energy of the exact axial solution. Hopf index three solitons are also the same as in the Skyrme-Faddeev and Nicole models, the $\tilde{\mathcal A}_{3,1}$ twisted axial configuration. It has an energy of $2.414$ when $V\=60000$. An axial $\mathcal A_{3,1}$ can be found as a saddle point of the minimisation if held by symmetry, with an energy of $2.476$ with a volume $V\=60000$. The axial solution held by symmetry is $\sim 1\%$ above the exact axial solution while the twisted $\tilde{\mathcal A}_{3,1}$ soliton is $\sim 1.5\%$ below the energy of the exact axial solution. The $\mathcal Q\=3$ soliton has an energy which is around $6\%$ above the conjectured energy bound $E\ge\mathcal Q^\frac{3}{4}$. 

As expected from (\ref{AFZaxial}) the minimal energy solution for Hopf index four in the AFZ model is an axial $\mathcal A_{2,2}$, see figure \ref{fig-afzmin}. This is the same as in the Skyrme-Faddeev model but not as in the Nicole model where the solution is an $\mathcal L^{1,1}_{1,1}$ link. The $\mathcal A_{2,2}$ solution has an energy of $2.852$ for $V\=130000$, $\lsim1\%$ above the expected energy from (\ref{AFZaxial}). Initial $\mathcal L^{1,1}_{1,1}$ configurations evolve towards an $\mathcal A_{2,2}$ type configuration but do not quite make it the whole way. The two strings do not seem to be able to unlink themselves and the result is a linked configuration but with the two unlinks lying almost on top of each other. 

At Hopf index five the axial $\mathcal A_{5,1}$ exact solution (\ref{AFZaxial}) is $\sim 15\%$ above the $E\ge\mathcal Q^\frac{3}{4}$ conjectured energy bound. Evaluating the AFZ energy of the static configuration in the Nicole model gives a much lower result than that of the exact axial solution at $\mathcal Q\=5$. Volume preserving minimisation using the symmetry breaking term results in two local minima with energies much lower than the $\mathcal A_{5,1}$ axial solution. The lowest energy configuration is a $\mathcal K_{3,2}$ trefoil knot, with an energy of $3.339$ for a volume $V\=150000$. This is $\sim 1.5\%$ above the conjectured bound. The isosurface near to the soliton's position forms a band-like structure rather than the usual string-like structure seen with trefoil knots in the Nicole or Skyrme-Faddeev model, (see \cite{nicvp} and \cite{knots}). The other static solution is an $\mathcal L^{1,1}_{1,2}$ link, like the minimal energy solutions in the Nicole and Skyrme-Faddeev models. It has an energy of $3.451$ when $V\=190000$, which is $\sim 2\%$ higher than the knotted solution. The trefoil knot appears to be the minimal energy soliton at $\mathcal Q\=5$, but the small difference in energy and the fact that no configuration transitions are seen makes this proposition inconclusive. Axial $\mathcal A_{5,1}$ initial conditions tend to twist, similarly to the $\tilde{\mathcal A}_{3,1}$ configuration. An approximately axial $\mathcal A_{5,1}$ configuration with a volume $V\=225000$ has an energy of $3.924$ which is $\sim1.5\%$ above the expected value. This configuration is not static as it tends to twist itself until a section of the string pinches off and the solution collapses. Conducting minimisation on a lattice with axial symmetry may result in an axial configuration with an energy much closer to the exact solution. 

Hopf index six has a static solution that differs from the Nicole and Skyrme-Faddeev models. As the $\mathcal A_{3,2}$ exact axial solution (\ref{AFZaxial}) has an energy that is only $1\%$ above the conjectured bound it seems likely it will be the minimal energy solution. A static $\mathcal A_{3,2}$ configuration has an energy of $3.907$ for a volume $V\=160000$. This is $\lsim1\%$ above the energy of the exact axial solution, as was the $\mathcal A_{2,2}$ solution at Hopf index $\mathcal Q\=4$. A range of initial conditions can be used, where the resultant minimisation leads to an $\mathcal A_{3,2}$ type configuration, or something near to an $\mathcal A_{3,2}$. The same issue with configuration transition occurs at $\mathcal Q\=6$ as did at $\mathcal Q\=4$. The $\mathcal L^{1,1}_{2,2}$, $\mathcal L_{1,1}^{2,2}$, $\mathcal K_{3,2}$ and $\mathcal K_{4,3}$ initial configurations all move towards an $\mathcal A_{3,2}$ type configuration, but the position strings seem unable to reconfigure themselves into the correct formation. The result are knots or links where the strings all approximately in a plane, almost on top of each other. 

Hopf index seven has only one solution and it is the same type as in the Nicole and Skyrme-Faddeev models, a trefoil knot. The $\mathcal K_{3,2}$ has an energy of $4.418$ with $V\=180000$. It is $\sim17\%$ lower in energy than the exact axial solution (\ref{AFZaxial}) and is $\sim2.5\%$ above $\mathcal Q^\frac{3}{4}$. All possible starting configurations lead to a trefoil knot when $\mathcal Q\=7$, including $\mathcal L^{1,1}_{2,3}$, $\mathcal L_{1,2}^{2,2}$, $\mathcal K_{5,2}$ and $\mathcal K_{4,3}$ type initial configurations. Axial initial conditions twist once again under minimisation. In fact, they twist too quickly for the axial profile function to minimise and an axial energy to be found. This is likely to be the case for all prime $\mathcal Q>7$, where the minimal energy configurations are likely to be knots and links, as in the Nicole and Skyrme-Faddeev models. 

The Hopf index eight axial $\mathcal A_{4,2}$ exact solution (\ref{AFZaxial}) has an energy $3\%$ above $\mathcal Q^\frac{3}{4}$. The minimal energy solution found by volume preserving minimisation from a range of initial conditions is a twisted $\mathcal A_{4,2}$ type configuration, denoted $\tilde{\mathcal A}_{4,2}$. It has an energy of $4.954$ for $V\=130000$, around  $4\%$ above $\mathcal Q^\frac{3}{4}$. That makes it $\sim1\%$ above the exact axial energy (\ref{AFZaxial}). This would be in line with the energies found when $\mathcal Q\=4,6$, except this solution is twisted. The configuration seems to be robust, since several types of initial conditions lead to an $\tilde{\mathcal A}_{4,2}$ type configuration, including $\mathcal L^{1,1}_{3,3}$, $\mathcal L_{2,2}^{2,2}$, $\mathcal K_{5,2}$ and $\mathcal A_{4,2}$ type fields. As with the axial solutions for $\mathcal Q\=4$ and $6$, the minimisation of other knot and linked configurations results in configurations close to the $\tilde{\mathcal A}_{4,2}$ configuration, where the strings are not able to reconnect in the correct orientation under volume preserving minimisation with symmetry breaking. Imposing axial symmetry in the initial conditions can lead to an $\mathcal A_{4,2}$ axial configuration, which has an energy of $4.978$ when $V\=110000$. This is $\sim1.5\%$ above the exact axial solutions energy. As the energy of the twisted $\tilde{\mathcal A}_{4,2}$ configuration is approximately what might be expected for the energy of an axial configuration under volume preserving minimisation it is not clear whether this is the minimal energy $\mathcal Q\=8$ soliton configuration in the AFZ model or not. The twisting may be an artefact of the Nicole symmetry breaking term in the modified minimisation. The twisting may minimise the Nicole component to such an extent that breaking the symmetry kills the axial solution. Investigation using a larger domain, $\Omega$, may help shed light on this. Perhaps another method of breaking the symmetry could also be used to back up these results. In either case, the Hopf index eight soliton differs from the solitons in the Nicole and Skyrme-Faddeev models. 

\subsection{Increasing \mbox{\boldmath $\epsilon$}}

The results presented above were found using mainly $\epsilon\=0.01$. On occasion short runs using $\epsilon\=0.02-0.05$ may be performed to overcome breakdowns in the minimisation process resulting from symmetry problems in the AFZ model. Taking $\epsilon\ge0.05$ during minimisation can have a large influence on the resulting static solutions for some Hopf charges. Increasing $\epsilon$ for $\mathcal Q\=1,2,3,4,5$ and $7$ does not alter the topology of the solitons. The only change is in the energy, where the field has been slightly altered to minimise the increased proportion of Nicole model energy present in the minimisation process. 

At Hopf index $6$, minimising the knotted configurations with a larger $\epsilon$ leads to a $\mathcal K_{3,2}$ type configuration. For small $\epsilon$, they minimise towards a $\mathcal K_{3,2}$ where the position is squashed into a plane and the string sections are close together. For larger $\epsilon$ the strings move apart, but the knot still remains somewhat squashed. The linked configurations evolve towards an $\mathcal L^{1,1}_{2,2}$ type configuration when $\epsilon>0.05$. This might be expected since the minimal energy solution in the Nicole model is of this type. The $\mathcal A_{3,2}$ configuration can also be altered by increasing $\epsilon$. Taking $\epsilon$ between $0.05$ and $0.10$ can evolve the field towards a twisted $\tilde{\mathcal A}_{3,2}$ type configuration, but with increased AFZ energy. In all cases, the energy is significantly larger than the small $\epsilon$ minimisations.

Similar results occur at Hopf index $8$. For $\epsilon<10$ the evolution tends towards an $\tilde{\mathcal A}_{4,2}$ twisted configuration. If $\epsilon$ is increased further the configuration splits into an $\mathcal L_{2,2}^{2,2}$ type link configuration. An $\mathcal L^{1,1}_{3,3}$ type configuration evolves towards an $\tilde{\mathcal A}_{4,2}$ twisted configuration for low $\epsilon$, but the two strings do not seem to be able to cross over to the $\tilde{\mathcal A}_{4,2}$ configuration using the symmetry breaking minimisation. Without the symmetry breaking the minimisation breaks down. Increasing $\epsilon$ leads to an $\mathcal L^{1,1}_{3,3}$ type configuration like that seen in the Skyrme-Faddeev model, where they lie close to each other almost in a plane. This is different to the static $\mathcal L^{1,1}_{3,3}$ configuration in the Nicole model, which has two unlinks linked perpendicular to each other. Again these configurations have a larger AFZ energy than for small $\epsilon$. This behaviour might account for the broken symmetry minimisation resulting in an $\tilde{\mathcal A}_{4,2}$ twisted configuration rather than an axial $\mathcal A_{4,2}$ configuration for small $\epsilon$. 

\section{Solitons in Conformal Skyrme-Faddeev Models}

The Nicole and AFZ models have been investigated both analytically and numerically. The most recent numerical developments have been detailed in the previous two chapters. Any linear combination of these two models will also result in a scale invariant theory with topological solitons. The set of all linear combinations of the Nicole and AFZ model, where the coefficients sum to one, form a one parameter family of conformal field theories permitting Hopf solitons. They will be referred to as conformal Skyrme-Faddeev (CSF) models. 

Differences in the topology of minimal energy solutions at a given Hopf index in the Nicole and AFZ models has already been noted. This is clearest in the case of square charges, $\mathcal Q\=n^2$. Solitons with $\mathcal Q\=n^2$ in the AFZ model are axial with equal winding about the two torus angles. Solitons with $\mathcal Q\=n^2$ in the Nicole model are links and knots. Investigation of solitons in the conformal Skyrme-Faddeev models and the topological transition of solutions across the set could shed some light on the nature of the solutions of the Nicole and AFZ models. They may also lead to some insights about the Skyrme-Faddeev model and its solitons. 

The one parameter family of conformal Skyrme-Faddeev models, $\mathcal L(\theta)$, are 
\be
\mathcal L(\theta)=\cos^2(\theta)\mathcal L_{Ni}+\sin^2(\theta)\mathcal L_{AFZ}.
\ee
Both Lagrangian components have been previously introduced (\ref{LNi}), (\ref{LAFZ}), and $\theta\in[0,\frac{\pi}{2}]$. The two extremes of the set give each model individually, the Nicole model is recovered when $\theta\=0$ and the AFZ model is recovered when $\theta\=\frac{\pi}{2}$. The static energy of this set of models is
\be
E_\theta=\int \dfrac{\cos^2(\theta)\Lambda^\frac{3}{2} + \sin^2(\theta)H^\frac{3}{4}}{32\pi^2\sqrt{2}\cos^2(\theta)+16\pi^2 2^\frac{3}{4}\sin^2(\theta)}\,d^3x.
\ee
This energy has been normalised so that $E_0\=E_\frac{\pi}{2}\=1$ for $\mathcal Q\=1$, i.e.\ the energy of the simplest Hopf map is set to one in both the Nicole and AFZ models. 

\subsection{Volume Preserving Flow}

The CSF models are clearly scale invariant $\forall\theta$, so energy minimisation can be conducted using volume preserving flow. The energy densities are
\be
\mathcal E_\theta=\cos^2(\theta)\mathcal E_{Ni}+\sin^2(\theta)\mathcal E_{AFZ}+\lambda_\theta\lb1-\bphi\cdot\bphi\rb.
\ee
Each model will require its own Lagrange multiplier, $\lambda_\theta$ and the energy densities of the Nicole and AFZ models are as defined previously, i.e.\
\be
\mathcal E_{Ni} = \lb\d_i\bphi\cdot\d_i\bphi\rb^\frac{3}{2} \equiv \Lambda^\frac{3}{2},\qquad \mathcal E_{AFZ} = \left[\lb\d_i\bphi\times\d_j\bphi\rb\cdot\lb\d_i\bphi\times\d_j\bphi\rb\right]^\frac{3}{4} \equiv H^\frac{3}{4}.
\ee 
The equations of motion for this model are just a linear combination of the individual equations of motion, leading to gradient flow given by
\be
\bF_\theta=\cos^2(\theta)\bF_{Ni}+\sin^2(\theta)\bF_{AFZ}+\lambda_\theta\bphi.
\label{Fmix}
\ee
$\bF_{Ni}$, (\ref{FNi}), and $\bF_{AFZ}$, (\ref{FAFZ}), have previously been given. The Lagrange multiplier is found as the flow must map to solutions on the two-sphere, i.e.\ $\bphi\cdot\bF_\theta\=0$, such that
\be
\lambda_\theta=-\cos^2(\theta)\bphi\cdot\bF_{Ni}-\sin^2(\theta)\bphi\cdot\bF_{AFZ}.
\ee 
Volume preserving flow  can be applied to fix a scale for the energy minimisation. 

\subsection{Solitons and Transitions}

The transition of soliton energy and topology is investigated for Hopf index $\mathcal Q\=4$. Static configurations at either end of the spectrum differ topologically. This would also be true for all square Hopf charges, $\mathcal Q\=n^2$, with $n>1$ as well as $\mathcal Q\=5$, $6$ and possibly $8$ (see section \ref{AFZsection} and \cite{nicvp}). Hopf index $\mathcal Q\=4$ is the lowest charge for which this occurs. The main questions are:
\begin{itemize}
\item Do configurations alter slowly and smoothly as $\theta$ changes?
\item Do two local minima exist for some values of $\theta$?
\item Do two topologically different solutions have the same energy for some $\theta$?
\item Does an intermediate soliton configuration exist in between the two extreme $\theta$'s?
\item Does a sudden topological change in solutions occur at a given $\theta$?
\end{itemize}

\subsubsection{Variation of \mbox{\boldmath $\theta$}}

\begin{table}[h!]
\vspace{-20pt}
 \begin{center}
\begin{tabular}{  | c | c | c | c | c | c | c | c | c | c | }
\hline
\multicolumn{9}{|c|}{\large{\textbf{CSF Energy for a Range of \mbox{\boldmath $\theta$} \T \B}}} \\
\hline	
\multicolumn{4}{|c|}{\textbf{$\mathcal L^{1,1}_{1,1}$ link \T \B}} &  & \multicolumn{4}{|c|}{\textbf{Axial $\mathcal A_{2,2}$}} \\ 
\hline	
$\theta$ \T \B & $\sin^2(\theta)$ & $E$  &  $E/\mathcal Q^\frac{3}{4}$  & &  $\theta$ & $\sin^2(\theta)$ & $E$  &  $E/\mathcal Q^\frac{3}{4}$\\ \hline\hline
0.00	&	0.000	&	3.165	&	1.119	&	&	0.80	&	0.515	&	3.145	&	1.112	\\ \hline
0.05	&	0.002	&	3.165	&	1.119	&	&	0.81	&	0.525	&	3.143	&	1.111	\\ \hline
0.10	&	0.010	&	3.165	&	1.119	&	&	0.82	&	0.535	&	3.141	&	1.110	\\ \hline
0.15	&	0.022	&	3.165	&	1.119	&	&	0.83	&	0.545	&	3.139	&	1.110	\\ \hline
0.20	&	0.039	&	3.164	&	1.119	&	&	0.84	&	0.554	&	3.136	&	1.109	\\ \hline
0.25	&	0.061	&	3.164	&	1.118	&	&	0.85	&	0.564	&	3.134	&	1.108	\\ \hline
0.30	&	0.087	&	3.163	&	1.118	&	&	0.86	&	0.574	&	3.132	&	1.107	\\ \hline
0.35	&	0.118	&	3.162	&	1.118	&	&	0.87	&	0.584	&	3.129	&	1.106	\\ \hline
0.40	&	0.152	&	3.160	&	1.117	&	&	0.88	&	0.594	&	3.127	&	1.105	\\ \hline
0.45	&	0.189	&	3.159	&	1.117	&	&	0.89	&	0.604	&	3.124	&	1.104	\\ \hline
0.50	&	0.230	&	3.157	&	1.116	&	&	0.90	&	0.614	&	3.121	&	1.104	\\ \hline
0.55	&	0.273	&	3.155	&	1.116	&	&	0.91	&	0.623	&	3.119	&	1.103	\\ \hline
0.60	&	0.319	&	3.153	&	1.115	&	&	0.92	&	0.633	&	3.116	&	1.102	\\ \hline
0.65	&	0.366	&	3.150	&	1.114	&	&	0.93	&	0.643	&	3.113	&	1.101	\\ \hline
0.70	&	0.415	&	3.147	&	1.113	&	&	0.94	&	0.652	&	3.110	&	1.100	\\ \hline
0.75	&	0.465	&	3.143	&	1.111	&	&	0.95	&	0.662	&	3.107	&	1.098	\\ \hline
0.79	&	0.500	&	3.139	&	1.110	&	&	1.00	&	0.708	&	3.090	&	1.092	\\ \hline
0.80	&	0.515	&	3.138	&	1.109	&	&	1.05	&	0.752	&	3.071	&	1.086	\\ \hline
0.81	&	0.525	&	3.137	&	1.109	&	&	1.10	&	0.794	&	3.049	&	1.078	\\ \hline
0.82	&	0.535	&	3.136	&	1.109	&	&	1.15	&	0.833	&	3.025	&	1.070	\\ \hline
0.83	&	0.545	&	3.134	&	1.108	&	&	1.20	&	0.869	&	2.999	&	1.060	\\ \hline
0.84	&	0.554	&	3.133	&	1.108	&	&	1.25	&	0.901	&	2.971	&	1.050	\\ \hline
0.85	&	0.564	&	3.132	&	1.107	&	&	1.30	&	0.928	&	2.943	&	1.040	\\ \hline
0.86	&	0.574	&	3.130	&	1.107	&	&	1.35	&	0.952	&	2.911	&	1.029	\\ \hline
0.87	&	0.584	&	3.129	&	1.106	&	&	1.40	&	0.971	&	2.884	&	1.020	\\ \hline
0.88	&	0.594	&	3.128	&	1.106	&	&	1.42	&	0.977	&	2.874	&	1.016	\\ \hline
0.89	&	0.604	&	3.127	&	1.105	&	&	1.45	&	0.985	&	2.858	&	1.011	\\ \hline
0.90	&	0.614	&	3.126	&	1.105	&	&	1.47	&	0.990	&	2.850	&	1.008	\\ \hline
0.91	&	0.623	&	3.125	&	1.105	&	&	1.50	&	0.995	&	2.839	&	1.004	\\ \hline
0.92	&	0.633	&	3.124	&	1.104	&	&	1.52	&	0.997	&	2.834	&	1.002	\\ \hline
0.93	&	0.643	&	3.122	&	1.104	&	&	1.54	&	0.999	&	2.830	&	1.001	\\ \hline
0.94	&	0.652	&	3.121	&	1.103	&	&	1.56	&	1.000	&	2.829	&	1.000	\\ \hline
0.95	&	0.662	&	3.120	&	1.103	&	&	1.57	&	1.000	&	2.828	&	1.000	\\ 
    \hline
  \end{tabular}
\end{center}
\vspace{-10pt}
\caption{\footnotesize{Static CSF energy for (left) $\mathcal L^{1,1}_{1,1}$ (right) $\mathcal A_{2,2}$ type initial conditions. $\mathcal L^{1,1}_{1,1}$ configurations with $\theta\gsim 0.87$ move towards an $\mathcal A_{2,2}$ type configuration. $\mathcal A_{2,2}$ energies with $\theta\lsim 0.87$ are the energies of the axial configurations, which are saddle points of the minimisation. $\mathcal A_{2,2}$ configurations with $\theta\ll 0.87$ do not have static solutions as the field will attempt to unwind to a single string and collapse. Axial configurations for $\theta$ just under $0.87$ can re-link and evolve towards the $\mathcal L^{1,1}_{1,1}$ minimal energy configuration. This transformation takes a very long time, due to the shallow gradient of the restricted energy functional in this region. Continued minimisation of the $\mathcal L^{1,1}_{1,1}$ solutions for $\theta>0.87$ leads towards near-axial configurations, with an energy slightly larger than that of the $\mathcal A_{2,2}$ minimal energy solutions.}}
\label{thetatable}
\vspace{-20pt}
\end{table}

The Nicole model soliton, at $\mathcal Q\=4$, is an $\mathcal L^{1,1}_{1,1}$ type linked configuration, with no other stable local minima existing \cite{nicvp}. In the AFZ model, it has been seen earlier that the soliton configuration representing the global energy minimum is an $\mathcal A_{2,2}$ type field configuration. 

The numerical calculations are conducted using the same details as in section \ref{AFZsection}. All solutions will have a volume $V\=160000$, resulting from initial configurations determined by rational maps $\mathcal A_{2,2}$ and $\mathcal L^{1,1}_{1,1}$ detailed previously. Minimisation where $\theta\approx\frac{\pi}{2}$ will require a much reduced timestep to circumvent issues relating to the AFZ model discussed in section \ref{AFZsection}. 

\begin{figure}[t!]
\centering
\vspace{-20pt}
\includegraphics[scale=0.6]{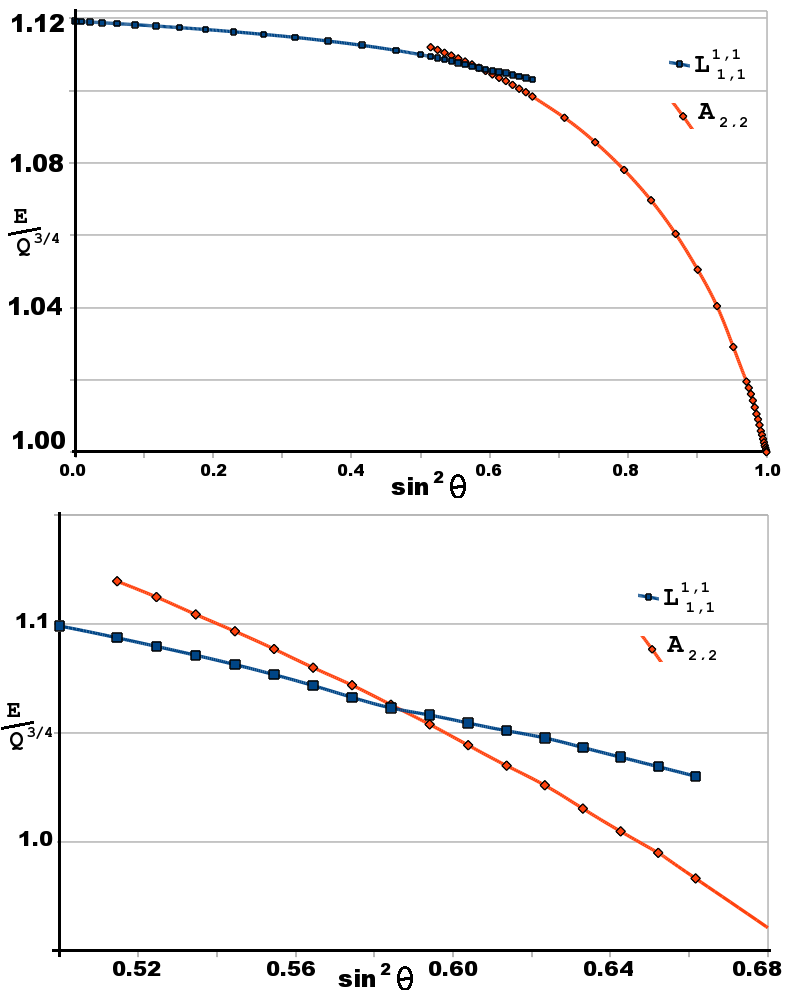}
\vspace{-10pt}
\caption{\footnotesize{Plots of the CSF energies for both configurations for a range of $\theta$'s, details found in table \ref{thetatable}. (top) Full range of $\theta$ and (bottom) region near transition $\mathcal L^{1,1}_{1,1}\leftrightarrow\mathcal A_{2,2}$. Plots are of $E/\mathcal Q^\frac{3}{4}$ against $\sin^2 \theta$, the coefficient of the AFZ term.}}
\label{fig-thetagraph}
\vspace{-10pt}
\end{figure} 

\begin{figure}[h!]
\centering
\vspace{-20pt}
\includegraphics[scale=1.3]{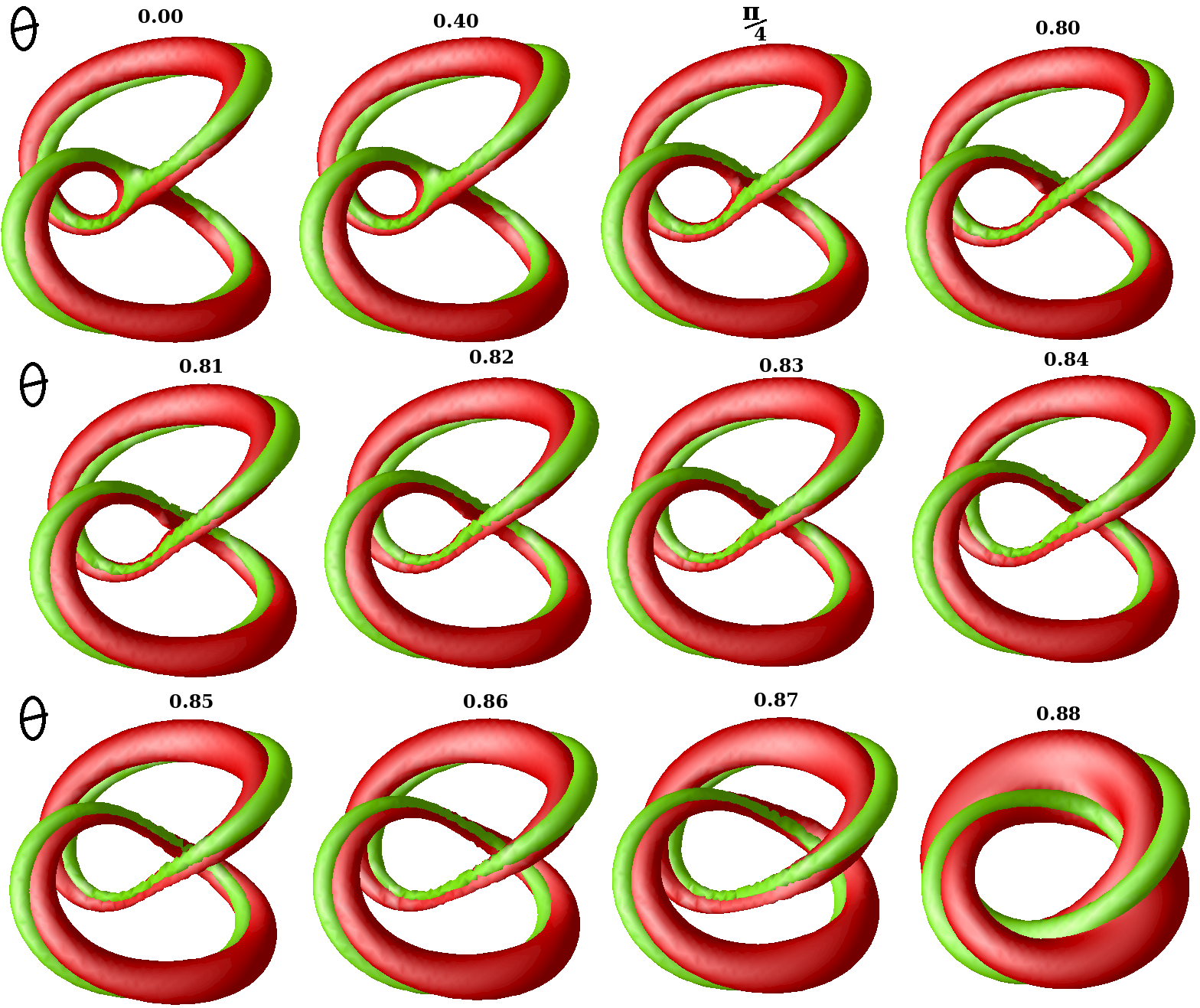}
\caption{\footnotesize{Position (red) and linking (green) plots for $\mathcal L^{1,1}_{1,1}$ minima for a range of $\theta$. Above $\theta=0.87$ solutions move towards $\mathcal A_{2,2}$ type configurations. For $\theta\gg\pi/4$, $\mathcal L^{1,1}_{1,1}$ initial conditions will lead to a similar configuration as with $\theta=0.88$ and have a slightly larger energy than an $\mathcal A_{2,2}$ minima for the same $\theta$.}}
\label{fig-link}
\includegraphics[scale=1.6]{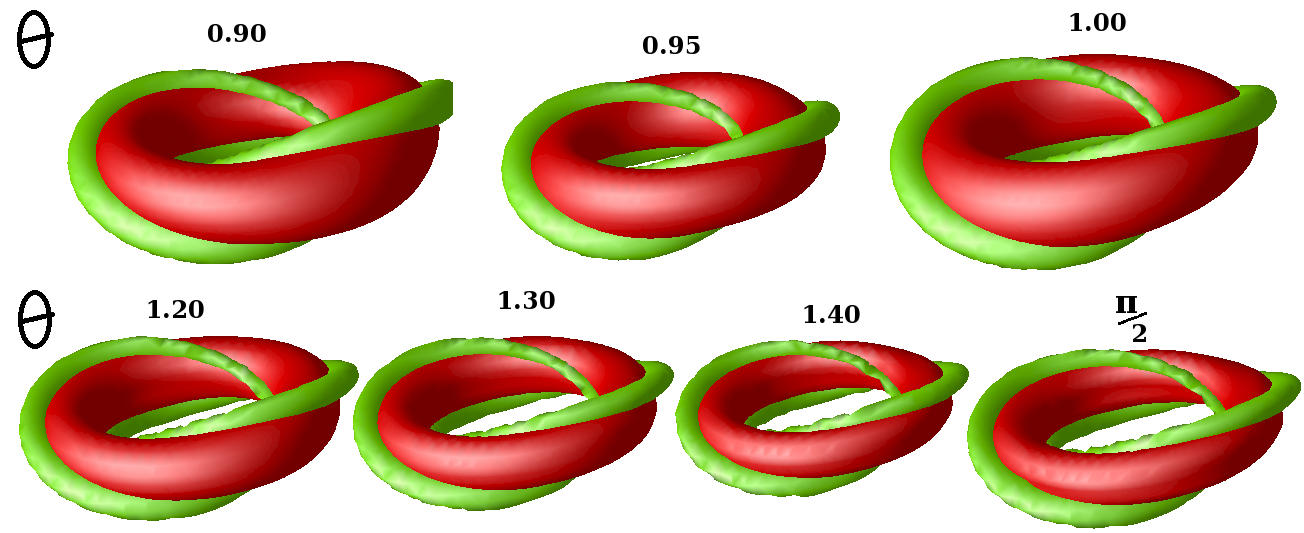}
\vspace{-10pt}
\caption{\footnotesize{Position (red) and linking (green) plots for $\mathcal A_{2,2}$ minima for a range of $\theta$. Configurations with $\theta\approx 0.9$ fluctuate between an $\mathcal A_{2,2}$ state like $\theta=1.00$ and a slightly linked state like figure \ref{fig-link} ($\theta=0.88$) with very little increase/decrease in energy. $\mathcal A_{2,2}$ configurations with $\theta\lsim 0.86$ tend to try to unwind to a highly twisted $\tilde{\mathcal A}_{4,1}\to\mathcal L^{1,1}_{1,1}$  and collapse along the way as a section becomes thin.}}
\label{fig-axial}
\vspace{-40pt}
\end{figure}

Table \ref{thetatable} shows a breakdown of the energies over a range of $\theta$ for both the link and the axial solitons. Figure \ref{fig-thetagraph} plots these results for the full range of $\theta$ (top) and in the range of topological change (bottom). For Hopf index $4$, $\mathcal L^{1,1}_{1,1}$ type configurations exist and are very similar in appearance when $\theta\lsim\frac{\pi}{4}$, see figure \ref{fig-link}. Axial $\mathcal A_{2,2}$ solitons exist and are similar in appearance for $\theta\gsim\frac{\pi}{3}$, see figure \ref{fig-axial}. The spacing between models considered, $\delta\theta$, is decreased for $\theta>\!\frac{3\pi}{7}$ due to issues relating to the AFZ model, as discussed in section \ref{AFZsection}. The range  $\theta\in\lb\frac{\pi}{4},\frac{\pi}{3}\rb$ is somewhat more complicated and more models are considered in this region. Selected results are plotted as isosurfaces near the position and linking in figures \ref{fig-link} and \ref{fig-axial}.   

The two components are self intersecting in the Nicole model and as $\theta\to\frac{\pi}{4}$ the self intersection is broken. For $\frac{\pi}{4}<\theta\lsim 0.87$ the unlinks move inward, eventually towards an $\mathcal A_{2,2}$ type configuration when $\theta\gsim 0.88$. Transitions between a configuration like figure \ref{fig-link} ($\theta\=0.81$) and the minimal configuration for a given $\theta$ in this range occurs over a very small energy range. For $\theta\=0.80$, the axial solution is only $0.2\%$ above the linked solution. This reduces quickly as $\theta\to 0.87$. Axial solutions lie in definite minima for $\theta\gsim\frac{\pi}{3}$. In the range $0.87\gsim\theta\gsim\frac{\pi}{3}$, solutions fluctuate between an axial configuration and a configuration similar to figure \ref{fig-link} ($\theta\=0.88$) during minimisation. The energy fluctuates up and down by a fraction of a percent. This indicates that either there are two local minima separated by a very low energy barrier, and the minimisation cannot properly distinguish them in this range, or that only one shallow local minima exists. Perturbations about this minima would then have a very similar energy if the gradient of the energy functional is small for a wider range of field configurations.

Clearly the topological transition of solutions for $\mathcal Q\=4$ occurs for $\theta>\frac{\pi}{4}$. This can be easily explained by analysing the simplest Hopf map for the Nicole and AFZ models in the context of the CSF models. The $\mathcal Q\=1$ energy in the Nicole model is $32\pi^2\sqrt{2}\approx467$ and in the AFZ model it is $16\pi^22^\frac{1}{4}\approx266$ when they are unscaled. This means that for the rescaling of the CSF models, each term gives an equally weighted contribution when $\theta\approx0.91$. Minimisation effects would have an equal contribution from both terms for the Hopf index one solution here. This matches with the results for the transition seen. One might expect the transition of the $\mathcal Q\=4$ soliton to occur slightly later than is found here. In the AFZ model $\mathcal Q\=4$ is a square charge and so the solution is expected to obtain the energy bound $16\pi^22^\frac{1}{4}\mathcal Q^\frac{3}{4}$ whereas the $\mathcal Q>2$ solitons are around $11\%$ above the energy bound $32\pi^2\sqrt{2}\mathcal Q^\frac{3}{4}$. Note that these bounds are not proven, they are simply the largest possible bound for behaviour like
\be
E\ge c\mathcal Q^\frac{3}{4},
\label{Eboundtheta}
\ee
where the constant $c$ is the energy of the simplest Hopf map in each case. These are the largest possible values of $c$ allowed. One might then expect a topological transition to occur at around $\theta\sim0.94$ rather than the $\theta\sim0.87$ seen. The cause of the fluctuations in this region is the most likely explanation for this.  

\section{Conclusion}

Non-axial solitons were found to be static solutions of the AFZ model using numerical simulations utilising a modified volume preserving flow. It has been shown that all solutions for $\mathcal Q<9$ are axial, twisted axial or knotted solutions. This does not rule out the existence of linked configurations at higher Hopf charges. The solutions show both similarities and differences to the Nicole and Skyrme-Faddeev models. Solutions clearly follow close to a potential lower energy bound of $E_{AFZ}\ge\mathcal Q^\frac{3}{4}$. Proof of this conjecture for both the AFZ and Nicole models would imply a universal feature of Hopf solitons, further to the solutions being knots and (un)links. Technical difficulties required the use of a symmetry breaking term to be added to the minimisation which slightly altered the problem. This symmetry breaking term was kept small and so the results should give a good reflection of the static solutions of the AFZ model. Further numerical or analytical evidence would be useful to back up the work undertaken here. 

The symmetry breaking minimisation motivates the investigation of a one parameter family of conformal Skyrme-Faddeev models. The two extremal models are the Nicole and AFZ models, with linear combinations of the two models completing the set. The family of models has been investigated for a Hopf index with qualitatively different topology of static solutions in the extremal models. The energy transition as the parameter is varied for these models has been found. Details regarding the topological transition of the static solutions has been investigated throughout the set of models. There is a smooth change in configurations over a short range of $\theta$. It is not clear whether two local minima exist for some $\theta$ or not. It is most likely that only one local minima exists for each $\theta$ but that these minima become shallow over a small range. The two qualitatively different configurations have very similar energies. In fact, the linked configurations look very similar to the axial configurations in the transition range. The solution transition is between axial and linked configurations. No intermediate configurations exist. The Nicole model contributes more energy than the AFZ model for a given coefficient and more than half the models have static solutions that are links. In the region where contributions from the Nicole term and the AFZ term are roughly equal, static solutions are much harder to find. Volume preserving flow finds these minima reasonably well for most models. For parameters near to the topological transition of static solutions, solitons are found to be slightly unstable up to small perturbations. These solutions appear to have near-zero modes associated with the breaking of axial symmetry, at least in the $\mathcal Q\=4$ case. A number of issues relating to CSF models are left open to investigation. An explicit analysis of the stability of solutions to such models would be of use. Do the equations of motion of such models simplify in the same way as with the Nicole and AFZ models and can an energy bound of the form (\ref{Eboundtheta}) be proven? 

These results provide further evidence of some universal features of Hopf solitons such as the appearance of knotted and linked solutions. Studying models, such as the Nicole and AFZ models, can help to explain the structures found in more physical models such as the Skyrme-Faddeev model. Volume preserving flow could be used to find static solutions to a range scale invariant field theories and might also be useful in the numerical construction instantons.

\section*{Acknowledgements}
I would like to thank the EPSRC for a research studentship, and Prof.\ P.\ M.\ Sutcliffe and D.\ Harland for useful discussions.

\end{document}